\documentclass[referee]{aa} 
\usepackage{graphicx}
\usepackage{natbib}

\usepackage{latexsym}
\usepackage{amsmath}
\usepackage{amssymb}
\usepackage{amstext}
\usepackage{color}

\def\cm3{cm$^{-3}$}

\def\3{$^{13}$CO}
\def\2{$^{12}$CO}

\usepackage{txfonts}
\begin{document}

\title{High-resolution radio study of SNR~IC~443 at low radio frequencies}

\author{G. Castelletti\inst{1,2}\footnote[1]{Member of the Carrera del 
Investigador Cient\'\i fico of CONICET, Argentina}
\and G. Dubner\inst{1}\footnotemark[1] 
\and T. Clarke\inst{3} 
\and N.~E. Kassim\inst{3}
}

\offprints{G. Castelletti}

   \institute{Instituto de Astronom\'\i a y F\'\i sica del Espacio (IAFE, CONICET-UBA),
              CC67, Suc.28, 1428, Buenos Aires, Argentina. \\                  
              \email{gcastell@iafe.uba.ar}
\and
             Facultad de Ciencias Exactas y Naturales (Universidad de Buenos Aires). 
\and
             Remote Sensing Division, Code 7213, Naval Research Laboratory, 
4555 Overlook Avenue, SW, Washington DC, USA.\\
             }

   \date{Received <date>; Accepted <date>}

 
  \abstract
   {}
   {To investigate in detail the morphology at low radio frequencies of the supernova 
remnant (SNR)~IC~443 and to accurately establish its radio continuum spectral properties.
} 
{We used the VLA in multiple configurations to produce  
high resolution radio images of  
IC~443 at 74 and 330~MHz. From these data we 
produced the first sensitive, spatially resolved, spectral analysis of the radio 
emission at long wavelengths. The changes with position in the radio spectral index 
were correlated with data in near infrared (NIR) from 2MASS, in gamma-rays from 
\it VERITAS\rm, and with the molecular $^{12}$CO~($J$=1--0) line emission.
}
{The new image at 74~MHz has \it HPBW \rm= 35$^{\prime\prime}$, rms=30~mJy~beam$^{-1}$ 
and at 330~MHz \it HPBW \rm= 17$^{\prime\prime}$ and rms=1.7~mJy~beam$^{-1}$. 
The integrated flux densities for the whole SNR are 
$S_{\mathrm{74\,MHz}}^{\mathrm{SNR}}=470\pm51$~Jy and 
$S_{\mathrm{330\,MHz}}^{\mathrm{SNR}}=248\pm15$~Jy. 
Improved estimates of the integrated spectrum were derived taking into account a 
turnover to fit the lowest frequency measurements in the literature. Combining our
measurements with existing data, we derive an integrated spectral index 
$\alpha_{\mathrm{10\,MHz}}^{\mathrm{10700\,MHz}}=-0.39\pm0.01$
with a free-free continuum optical depth at 330~MHz 
$\tau_{330}\sim 7 \times 10^{-4}$ ($\tau_{10}=1.07$); all measurements above $\sim10$~MHz 
are equally consistent with a power law spectrum.
For the pulsar wind nebula associated with the 
compact source CXOU~J061705.3+222127, we calculated 
$S_{\mathrm{330\,MHz}}^{\mathrm{PWN}}=0.23\pm0.05$~Jy, 
$S_{\mathrm{1420\,MHz}}^{\mathrm{PWN}}=0.20\pm0.04$~Jy, and
$\alpha_{\mathrm{330\,MHz}}^{\mathrm{8460\,MHz}}\sim 0.0$.
Substantial variations are observed in spectral index between 74 and
330~MHz across IC~443. The flattest spectral components ($-0.25\leq\, \alpha \,\leq-0.05$) 
coincide with the brightest parts of the SNR along the eastern border, with
an impressive agreement with ionic lines as observed in the 2MASS \it J \rm 
and \it H \rm bands. The diffuse interior of IC~443 has a 
spectrum steeper than found anywhere in the SNR 
($-0.85\leq\, \alpha \,\leq-0.6$), while the southern ridge again has a flatter spectrum
($\alpha \sim -0.4$). At the available statistics the \it VERITAS \rm $\gamma$-ray emission 
strikingly matches  
the CO distribution, but no clear evidence is found for a morphological correlation between 
the TeV distribution and radio emission. 
}
{The excellent correspondence between the eastern radio flattest spectrum region and 
NIR ionic lines strongly suggests that the passage of a fast
dissociating J-type shock across the interacting molecular cloud dissociated the 
molecules and ionized the gas. We therefore conclude that thermal absorption at 74~MHz
($\tau_{74}$ up to $\sim0.3$) is responsible for the localized spectral index 
flattening observed along the eastern border of IC~443. Towards the interior of IC~443 the spectrum
is consistent with those expected from linear diffusive shock acceleration, while the flatter spectrum
in the southern ridge is a consequence of the strong shock/molecular cloud interaction.}

\titlerunning{High-resolution radio study of SNR~IC~443 at low radio frequencies}
\authorrunning{\textsc{Castelletti et al.}}

\maketitle
%

\section{Introduction}
The importance of radio continuum studies of SNRs for understanding shock acceleration processes 
\citep{rey92,and96} as well as intrinsic and extrinsic interactions with ionized gas and the
interstellar medium (ISM) \citep{dul75,kas89} has been long appreciated. However the magnitude
of the observable effects are often subtle, requiring a large leverage arm in frequency space
to discern. The lack, until recently, of sufficient angular resolution and sensitivity at the
lowest frequencies has made progress in such studies slow. The advent of high resolution, low 
frequency observations with instruments like the VLA and GMRT are now changing this picture. 
In particular, VLA sub-arcminute resolution imaging below 100~MHz has been important 
for discerning spatially resolved intrinsic and extrinsic thermal absorption. 
Examples now include unshocked
ejecta inside Cas~A \citep{kas95}, thermal filaments within the Crab nebula \citep{bie97}, and
ionized gas in the ISM along the line of sight towards W49B \citep{lac01}. Later on, 
\citet{bro05} spatially resolved the ionized boundary marking the SNR/molecular cloud (MC) 
interface in 3C~391, suggesting such signatures could be both common and important for 
delineating elusive SNR/MC interactions.
Most recently, \citet{cas07} detected strong 74~MHz free-free absorption at the
interface between the SNR~W44 and the photo dissociation region of a neighbouring HII region.

In this paper we extend such high resolution, low frequency radio studies to the classic 
SNR~IC~443, one of the clearest a priori known cases of a remnant interacting with its 
cloudy surroundings. We present the first, high resolution radio studies of this object, 
and interpret our results in context of recent IR and high energy observations. IC~443 is 
a particularly attractive SNR for low frequency
studies both because of its large angular size and because its outer Galaxy location leave it 
relatively well isolated from the confusing effects normally contaminating studies towards inner
Galactic SNR/MC complexes.

\subsection{An overview of SNR~IC~443}
Observed in the radio domain \object{IC~443} (G189.1+3.0) is among the larger SNRs in angular size 
catalogued in our Galaxy.
It consists of two connected, roughly spherical, shells of radio
synchrotron emission, which are centered at different locations 
\citep[the names ``shell A'' and ``shell B'' are sometimes used to refer their locations 
towards the east and west halves of the remnant,][]{bra86a}.
On its eastern side, IC~443 has a rim-brightened morphology, while towards the western 
half the surface brightness is dimmer. 
A third incomplete and faint shell 
\citep[called by][``shell C'']{bra86a} is also evident 
extending beyond the northeast periphery of the remnant when the radio images are 
displayed with very high contrast. 
Based on both its morphology and soft X-ray spectrum, ``shell C'' was proposed to 
be a different SNR called G189.6+3.3, that overlaps with IC~443 \citep{asa94}.
At present it is still questionable if it is a SNR at all, and, if it is, then
it is unclear whether \object{G189.6+3.3} is
interacting with IC~443 \citep{asa94,lea04,lee08}. 

At optical wavelengths the appearance of IC~443 resembles that observed in the 
radio band. Spectroscopic and photometric observations of this remnant reveal a 
complex filamentary structure varying in brightness and shape, which is mainly 
composed of Balmer-line emission, strong
[CaII], [SII], [NII], [OIII], as well as faint [FeII] and [FeIII] emission regions
\citep{fes80}.  
The emission-line ratios are consistent with those of a well-evolved SNR \citep{fes80}.

Kinematical considerations based on optical systemic velocities place the SNR 
between 0.7 and 1.5~kpc \citep{loz81}, although  
under the assumption of a physical relationship 
of IC~443 with the nearby HII region Sharpless~249, it might be located slightly 
more distant between 1.5 and 2~kpc \citep{fes84}.
A distance of 1.5~kpc is adopted in most of the works related to IC~443.

Different X-ray observations carried out with the \it Einstein 
Observatory \rm \citep{pet88}, \it Ginga \rm \citep{wan92}, 
\it ROSAT \rm \citep{asa94}, \it ASCA \rm \citep{kaw02}, 
and \it XMM-Newton \rm \citep{tro06} show 
a bulk of thermal centrally peaked X-ray emission within the radio rim. 
On the basis of the radio/X-ray morphology and the X-ray properties, several
authors proposed that IC~443 belongs to the class of ``mixed-morphology'' or 
``thermal-composite'' SNRs \citep{rho98}.

The compact X-ray source \object{CXOU~J061705.3+222127} is thought to be the remnant of the
explosion that gave rise to IC~443.
This compact relic, discovered using data 
at high radio frequencies and in X-rays,
is embedded within the remnant and is located close to its southern edge \citep{olb01}.
Albeit no pulsations were detected from the neutron star, the measure of its transverse 
velocity provides an estimate for the age of both the SNR and its associated neutron 
star, of about $3\times10^{4}$~yr \citep{olb01}. An alternative estimate 
of $\sim$3000~yr was previously proposed by \citet{pet88} based on a model of its 
X-ray emission.

IC~443 is a prototypical case of a SNR impacting dense interstellar molecular gas. 
The giant molecular cloud, as mapped by CO and HCO$^{+}$ observations, 
forms a ring in the foreground 
and appears to be interacting with the remnant at several locations from
north to south \citep{dic92}. This phenomenon, which seems to be responsible for the 
multiwavelength observational picture of the remnant, was unambiguously confirmed by the 
presence of several OH~(1720~MHz) masers as well as IR cooling lines from  H$_{2}$ 
\citep[][and references therein]{rho01,hew06}.
Additionally, HI observations show well-defined filamentary structures 
in the northeastern region of the remnant, where the brightest 
optical filaments were located \citep{lee08}. These features are interpreted as the 
recombined neutral gas in an atomic shock \citep{lee08}. 

Another interesting characteristic of IC~443
is the presence of associated high energy sources. 
The EGRET GeV source (3EG~J0617+2238) was detected overlapping the remnant \citep{esp96}. 
In addition, TeV
gamma-rays from the central region of the remnant, coincident with the maser 
emission, were recently 
reported from observations carried out with MAGIC, \it VERITAS\rm, and 
\it FERMI \rm telescopes \citep{alb07,acc09,abd10}. 
It has been suggested that the correlation of gamma rays with molecular gas arises from
the pion decay of hadronic cosmic rays generated by the interaction of the SNR shock with
dense molecular material \citep{hum08,abd10}.

\section{Observations and data reduction}
The Very Large Array (VLA)\footnote{The Very Large Array of the National Radio Astronomy 
Observatory is a facility of the National Science Foundation operated under cooperative 
agreement by Associated Universities, Inc.} was used in multiple configurations to observe 
the large SNR~IC~443 at 74 and 330~MHz. A summary of the observations carried out on 
various dates between 2005 and 2007 is given in Table~\ref{parameters}.

\begin{table*}
\renewcommand{\arraystretch}{1.0}
\centering
\vspace{0.26cm}
\caption{Observational summary}
\begin{tabular}{lcccccc}  \hline \hline
Program & Observing      & VLA &   Bandwidth & Integration & Synthesized & Largest detectable\\
& dates          & Config. &   (MHz) &    time (hr) & beam (arcsec)& structure (arcsec)\\
\hline
\multicolumn{7}{c}{74 MHz Parameters} \\
\hline
AG0697   & 2006 Apr 8, 9, 10  &  A   & 6.3 & 9 & 25$\times$20 & 800 \\
AG0735   & 2006 Oct 27  & C  & 6.3   & 8.35  & 228$\times$214 & 7500 \\
AG0697   & 2007 Oct 21  & AB   & 6.3 & 6.85  & 28$\times$30 & 1300 \\
\hline
\multicolumn{7}{c}{330 MHz Parameters} \\
\hline
AG0697 & 2005 Sep 16  & C   & 6.3 & 4.7  & 62$\times$60 & 1800 \\
AG0697  & 2006 Apr 8, 9, 10  & A   & 1.5 & 9  & 6$\times$6 & 170 \\
AG0735  & 2006 Oct 27 & C   & 1.5 & 8.35   & 57$\times$53 & 1800\\
AG0735  & 2007 Apr 15, 16  & D   & 6.3 & 5.2  & 238$\times$205 & 4200 \\
AG0697  & 2007 Oct 21  & AB   & 1.5 & 6.85  & 22$\times$17 & 500 \\ \hline
\label{parameters}
\end{tabular}
\end{table*}

In order to help with the radio frequency interference (RFI) excision and mitigate the 
effects of bandwidth smearing all the data were acquired in  
multi-channel continuum mode (64 and 16 channels per polarization at 74 and 330~MHz, 
respectively). At both, 74 and 330~MHz, we first performed a bandpass calibration on 
either \object{3C~405}~(Cygnus~A), \object{3C~147}~(0542+498) or \object{3C~274}~(Virgo~A) 
using publicly available 
models\footnote{See http://lwa.nrl.navy.mil/tutorial/ to obtain source models 
in FITS format.}.
The initial antenna-based gain and phase corrections at 74~MHz were estimated from 
observations of either 3C~405 or 3C~274.
For the 330~MHz data, observations of 3C~147 were sufficient for both, flux density and
initial phase calibration at this frequency. 
In all cases the absolute flux scale was set according 
to the \citet{baa77} scale.
The NRAO Astronomical Image Processing Software (AIPS) package was used to process 
all the observations.

Large fields of view are a general characteristic when observing at long 
radio wavelengths and it is necessary to avoid distortions introduced in the 
image caused by the non-coplanarity of the baselines of the array. 
To overcome this, we employed a pseudo-three-dimensional Fourier inversion 
as implemented in the AIPS task IMAGR, in which the primary beam area of 
the 74 and 330~MHz data, $11^{\circ}.5$ and $\sim 2^{\circ}.5$,
respectively, is divided into multiple partially overlapping 
facets \citep{cor92}. 
Furthermore, to compensate for time variable ionospheric phase effects we 
performed successive rounds of self-calibration and imaging to the data from
each configuration separately at 74 and 330~MHz. Angle-invariant self-calibration
(as implemented in the AIPS task CALIB) is generally inadequate for compensating for
ionospheric effects across a large field view, particularly for the 74~MHz B and
A configuration \citep{kas07}. An exception is for fields dominated by a bright
source at the field center, as in the case of IC~443.
The final calibrated visibility data from all of the observations listed in 
Table~\ref{parameters} for each frequency were then combined into a 
single \it uv \rm data set, after which a further amplitude and phase 
self-calibration was performed. 
For the concatenated 330~MHz data we employed a multi-scale CLEAN algorithm 
in AIPS,  with three different scales sizes. This process is efficient to 
make high resolution images that are also sensitive to extended structures. 
All the resulting facet images were stitched together into one large field 
using AIPS task FLATN to create a single final image with a 
synthesized beam of $17^{\prime\prime}.27 \times 15^{\prime\prime}.81$, 
PA=$-24^{\circ}$, which represents an order of magnitude improvement in 
angular resolution over the 330~MHz image of 
\citet{bra86a} and that of \citet{hew06}. 
The sensitivity achieved in our image after correcting for primary beam 
attenuation is 1.7~mJy~beam$^{-1}$.

Concerning the observations at 74~MHz, we noted that the extended radio emission 
in the field was more properly imaged using the capability incorporated in the 
task IMAGR to switch back and forth between the SDI Clean \citep{ste84} and the 
usual CLEAN deconvolution strategy, depending on the contrast between the 
brightest residual pixel and the bulk of residual pixels after each major cycle.
Following this procedure, the final resolution of the first image of IC~443 obtained 
at 74~MHz after combining the data sets described in Table~\ref{parameters} and 
including primary beam corrections is
$36^{\prime\prime}.50 \times 31^{\prime\prime}.76$, PA=$34^{\circ}.71$. 
The rms noise level in the 74~MHz image is 30~mJy~beam$^{-1}$.

Ionospheric refraction and self-calibration normally introduce arbitrary
position shifts on low frequency images. These shifts are readily corrected by 
registering against background small diameter sources whose positions are known
from higher frequency observations. We corrected these shifts by measuring several 
small-diameter sources with respect to their known positions from the NRAO VLA
Sky Survey (NVSS) \citep[the latter has an astrometric accuracy better 
than $1^{\prime\prime}$ in both R.A. and dec.,][]{con98}. 
We determined and corrected for a mean positional difference 
of $0^{\mathrm{s}}.25\pm0^{\mathrm{s}}.06$ in R.A. 
and  $5^{\prime\prime}.15\pm1^{\prime\prime}.00$ 
in dec. at 74~MHz and $0^{\mathrm{s}}.08\pm0^{\mathrm{s}}.04$ in R.A. and 
$3^{\prime\prime}.71\pm0^{\prime\prime}.42$ in dec. at 330~MHz. 

\section{Results}
\subsection{Low frequency VLA images of SNR~IC~443}
\begin{figure}[ht]
\centering
\includegraphics[width=8cm]{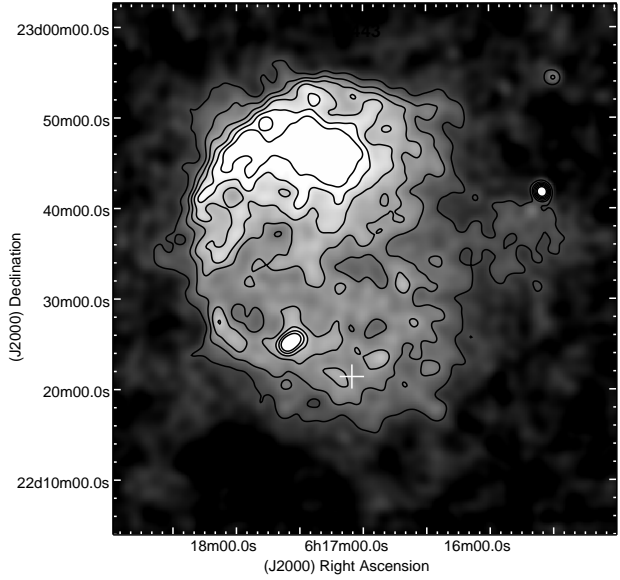}
\caption{Radio continuum image of IC~443 at 74~MHz.
This map has been corrected for the attenuation of the primary beam. 
The angular resolution of this image was smoothed to a beamsize of $50^{\prime\prime}$.
The grayscale is linear ranging from 25 to 250~mJy~beam$^{-1}$ and the contour levels
are 72, 120, 160, 200, 240, and 280~mJy~beam$^{-1}$. 
The plus symbol marks the position of the compact source CXOU~J061705.3+222127.
These observations provide the first 
subarcminute image of the remnant created at meter wavelengths.}
\label{74-subim}
\end{figure}

In Fig.~\ref{74-subim} we present a close-up view of the radio continuum 
emission at 74~MHz from IC~443. These observations provide the first 
subarcminute image of this remnant created at meter wavelengths. 
The data at 74~MHz are sensitive to smooth structures up to 
$\sim125^{\prime}$ in size, larger than the full extent of the SNR,
and thus are sensitive to the largest scale structure present in the SNR.

\begin{figure*}[ht]
\centering
\includegraphics[width=12cm]{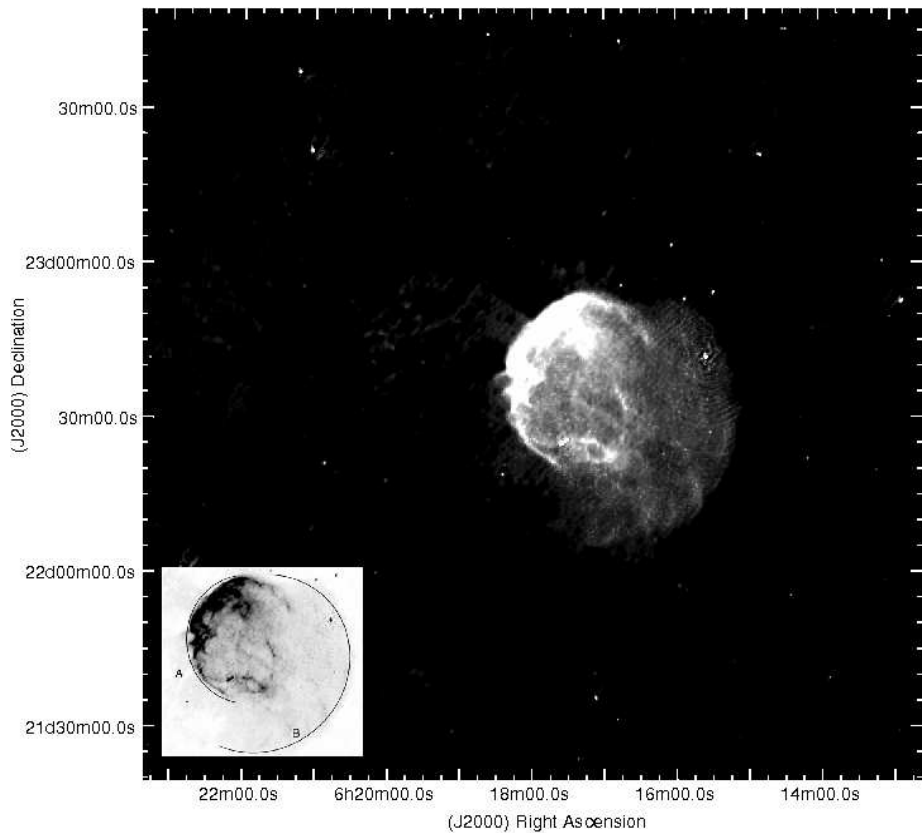}
\caption{The image shows the entire field of view of the VLA around the 
SNR~IC~443 at 330~MHz.
The image displayed   
includes primary beam correction. The synthesized beam is
 $17^{\prime\prime}\times16^{\prime\prime}$ with a position angle of $-24^{\circ}$, and the 
sensitivity level is 1.7~mJy~beam$^{-1}$.
This is the first high fidelity and high resolution view of the emission from 
IC~443 ever obtained at low radio frequencies. The inset with the
image of IC~443 at 330~MHz
is included to help in the location of the
bright and weak radio shells referred to in the literature as shells A and B, respectively.}
\label{330-fov}
\end{figure*}

\begin{figure*}[ht!]
\centering
\includegraphics[width=12cm]{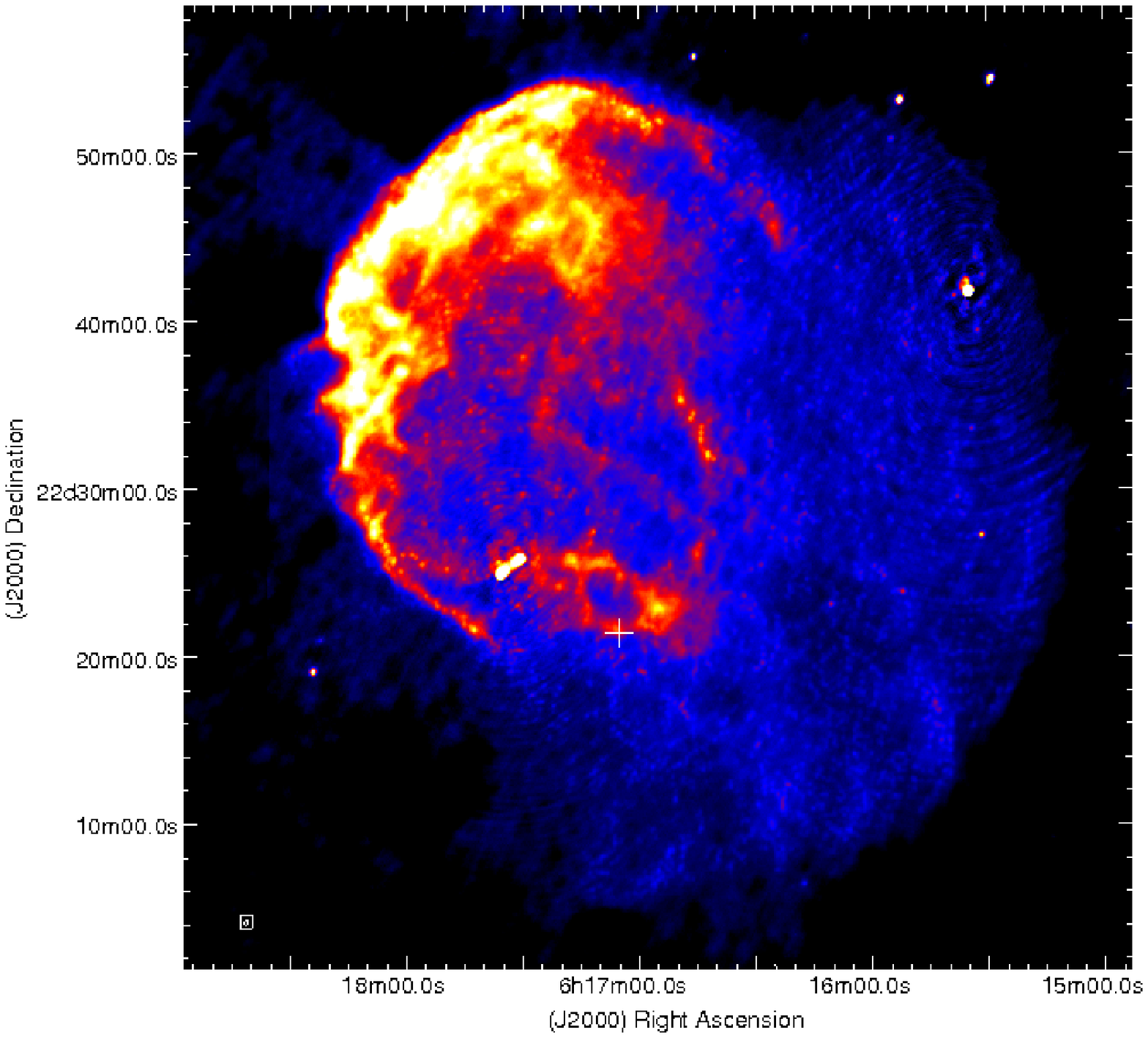}
\caption{A color image of the radio continuum emission from IC~443 at 330~MHz
constructed using multiple-configuration VLA observations.
The brightness range covered by the scale is between 3 and 35~mJy~beam$^{-1}$. 
The final beam size shown at bottom left is 
$17^{\prime\prime} \times 16^{\prime\prime}$ at a position angle of
$-24^{\circ}$. The noise level is 1.7~mJy~beam$^{-1}$ after primary beam 
correction. The color scale runs between 3 and 45~mJy~beam$^{-1}$.  
The white plus sign marks the location of the source CXOU~J061705.3+222127.}
\label{330-subim}
\end{figure*}

Figure~\ref{330-fov} displays with a resolution of $\sim17^{\prime\prime}$ the new 
VLA image of the $\sim2.5$ square degrees field of view centered on the remnant
observed at 330~MHz\footnote{The FITS files of the radio images of IC~443 at 74 and 330~MHz are available
online at http://www.iafe.uba.ar/snr/SNR.html.}.
An enlargement showing the detailed 330~MHz total intensity morphology of IC~443 is 
shown in Fig.~\ref{330-subim}.
The combination of data from the different VLA array configurations ensures that 
all scales of the radio emission are well represented in the 330~MHz image.

The distinctive morphological characteristics of IC~443, that comprise both the 
rim-brightened eastern side and the breakout into a region of significantly 
lower density to the west, were known from past radio observations.
For example, a lower resolution WSRT~327~MHz image \citep[][Fig.~5]{bra86a}
provides excellent surface brightness sensitivity to the most extended 
structures. Our 330~MHz image (Fig.~\ref{330-subim}) is quite complementary, with 
the higher angular resolution resolving for the first time at such low 
frequencies small scale structures covering almost the entire extent of the
eastern radio shell of the remnant, a region where previous observations have 
indicated the presence of ionic 
shocks as well as abundant shocked HI and optical filaments \citep{lee08}.
It is remarkable that towards the northeast and southeast ends of the 
eastern shell the brightest 
emission distribution becomes significantly narrower in appearance.
An average diameter of $\sim35^{\prime}$, or $\sim15$~pc at the assumed 
distance of 1.5~kpc, is measured in our map for the
bright radio shell in IC~443 (that, as mentioned in Sect.~1.1 
it is referred to in literature as shell A, and is depicted in the inset in Fig.~\ref{330-fov}). 
The low surface brightness emission gradually decreases outwards to the west
forming a more uniform radio shell of about $\sim52^{\prime}$ in diameter or $\sim23$~pc (the so 
called shell B, Fig.~\ref{330-fov}),
as measured on the new 330~MHz view of the remnant.

Additionally, the 330~MHz image of IC~443 serves to show that 
even at low radio frequencies part of the  
east outer border of the 
remnant is quite structured. It is particularly remarkable the indented 
morphology  of the bright rim at
R.A.$=06^{\mathrm{h}}\,18^{\mathrm{m}}\,15^{\mathrm{s}}$,
dec.$=+22^{\circ}\,37^{\prime}$, with tenuous radio
synchrotron emission detected ahead of the main shock. 
Such faint emission is seen as a weak and irregular radio halo that
in our 330~MHz image is notable at a mean level of $\sim 4\sigma$ 
extending up to $5^{\prime}$ ahead of the bright sharp boundary.
It is interesting to note that similar weak, diffuse emission
upstream of the main shock is also observed in the cases of Puppis~A
and W44, along the sides where the expanding blast wave encountered molecular
gas \citep[][and Castelletti et al. 2007]{cas06}.
A series of small protrusions are also evident along most of the extension of 
the eastern rim.  
The largest of these features located at
R.A.$=06^{\mathrm{h}}\,18^{\mathrm{m}}\,20^{\mathrm{s}}$,
dec.$=+22^{\circ}\,38^{\prime}$ emerges
about 2$^{\prime}$ radially from the border of the remnant. 
These structures can be also seen in the image of IC~443 at 1420~MHz 
presented by \citet{lee08}, who refer to them as ``spurs''.
The authors interpret that both, the halo and the ``spurs'', 
are likely originated from the interaction of the SNR with the 
surrounding inhomogeneous ambient medium, although an alternative explanation
based on a physical association between IC~443 and the SNR~G189.6+3.3 is not
entirely ruled out in their work. 
The low surface brightness emission from G189.6+3.3, positionally
coincident with IC~443, 
is clearly detected in our 330~MHz image extending outside the IC~443's eastern
shell\footnote{
The radio continuum emission from G189.6+3.3 is better represented in all its 
extension in the low resolution image at 327~MHz presented by \citet{bra86a}, 
displayed with the appropriate contrast to enhance it.}
near R.A.$=06{^\mathrm{h}}\,18{^\mathrm{m}}\,15{^\mathrm{s}}$,
dec.$=22^{\circ}\,46^{\prime}$. 

Towards the center of the SNR, in the southern half,
the most remarkable feature is a bright annular filament placed near
R.A.$=06^{\mathrm{h}}\,17^{\mathrm{m}}\,10^{\mathrm{s}}$,
dec.$= +22^{\circ}\,24^{\prime}$, which is immersed
in the faint and diffuse emission that dominates the inner part of IC~443. 
Such annular structure forms the central part of a more extended feature 
commonly referred in the literature as the southern sinuous ridge. The large 
southern ridge is defined below 
dec.$\sim +22^{\circ}\,35^{\prime}$ at 330~MHz and appears to 
be an extension of the southeast border of the remnant. Shocked CO gas with 
broad lines has been detected near this region of the SNR \citep{cor77,dic92}. 

Integrated flux densities estimates for IC~443 were made using the new 
observations, yielding S$_{\mathrm{74~MHz}}=470\pm51$~Jy and 
S$_{\mathrm{330~MHz}}=248\pm15$~Jy.
The quoted values have been corrected for the primary beam response and for the 
contribution of unrelated point sources overlapping the remnant.
The uncertainties in the measurements account for the statistical errors as well 
as the selection of integration boundaries.  

\subsection{The pulsar wind nebula}
The new sensitive image at 330~MHz offers, for the first time, a view of the 
low frequency counterpart of the low-luminosity plerionic nebula powered by the 
source CXOU~J061705.3+222127 observed both at higher radio frequencies with 
the VLA and in X-rays with  \it Chandra \rm and 
\it XMM-Newton \rm \citep{olb01,boc01,gae06}.

\begin{figure}[ht]
\centering
\includegraphics[width=8cm]{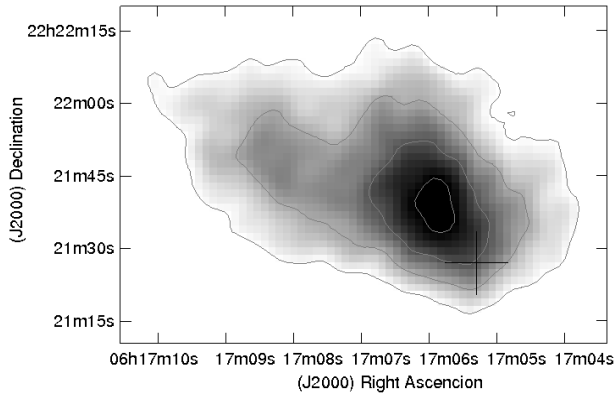}
\caption{A close-up image of the nebular emission around
CXOU~J061705.3+222127 at 330~MHz.
The position of CXOU~J061705.3+222127 is shown by the plus sign.
The grayscale varies from 19 to 25~mJy~beam$^{-1}$.
The contour levels on the image are traced at 19, 21, 23, and 25
mJy~beam$^{-1}$.}
\label{pwn}
\end{figure}

A close up image of the pulsar wind nebula (PWN) at 330~MHz is displayed in 
Fig.~\ref{pwn}. At 74~MHz the combined effect of the lower sensitivity, poorer 
angular resolution, and flatter intrinsic spectrum, conspire against its visibility 
at this low frequency.
As in X-rays, at low radio frequencies the source CXOU~J061705.3+222127
(marked in Fig.~\ref{pwn} with a plus sign) is placed at the apex of the 
nebula, although a radio counterpart for the point X-ray source is not 
detected at 330~MHz. 
The maximum embedded in the radio nebular emission,
with an intensity that peaks up to 25~mJy~beam$^{-1}$,
lies about  0$^{\prime}$.25 northeastward from
CXOU~J061705.3+222127. The cometary shaped  nebula extends along its major axis
approximately $1^{\prime}.25$ behind CXOU~J061705.3+222127 and $0^{\prime}$.21 ahead of it.
The integrated flux density over the entire nebula obtained from our image at
330~MHz is $S_{\mathrm{330\,MHz}}^{\mathrm{PWN}}=0.23\pm0.05$~Jy.
In addition we estimated the flux density of the PWN at 1420~MHz on 
the basis of the image of IC~443 presented by \citet{lee08} performed by 
combining VLA observations and single dish data taken from the Arecibo Telescope, 
obtaining $S_{\mathrm{1420\,MHz}}^{\mathrm{PWN}}=0.20\pm0.04$~Jy.

\subsection{Radio and optical emission of IC~443} 
\begin{figure*}[ht!]
\centering
\includegraphics[width=12cm]{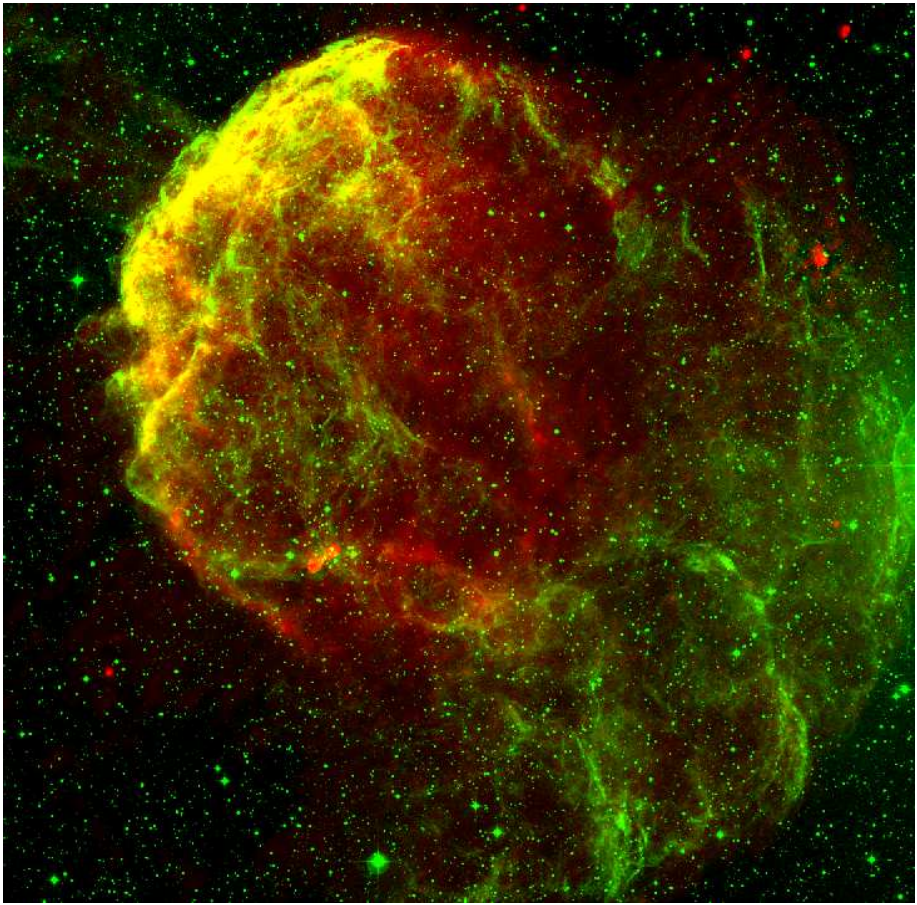}
\caption{A high resolution comparison between radio and optical emission from IC~443 SNR.
The green corresponds to optical emission from the Second Palomar Observatory
Sky Survey, while in red the 330~MHz radio emission is shown. The yellow regions 
are areas where emission in both spectral bands overlap.}
\label{optico-radio}
\end{figure*}

Figure~\ref{optico-radio} illustrates, with a spatial resolution better than
$20^{\prime\prime}$, the morphological comparison between 
the optical and radio emission from IC~443 obtained from the combination
in a false color image of our new VLA observations at
330~MHz (in red) with data from the Second Palomar Observatory Sky Survey
(in green). In this figure, features where both
spectral bands overlap are shown in yellow. 

The optical emission tracing the low density atomic gas reproduces the 
east-west asymmetry characteristic of the radio total intensity emission as 
well. The coincidence of the synchrotron-enhanced radio emission with
very strong optical filaments observed in H$_ {\alpha}$ and [SII] lines 
towards the east, suggests that the optical features delineate the position 
of cooling post-shock ISM gas.
Furthermore, near infrared emitting gas observed in this portion of the SNR
(see below Fig.~\ref{ir-radio}) together with abundant neutral gas, 
correlated both in space and velocity with the optical filaments, 
indicate the presence of radiative shocks propagating into gas of different densities.

The quality of the new 330~MHz image allows us to identify the close 
radio/optical correspondence in
most of the small-scale radio structures observed as extensions 
from the bright eastern portion of the shell. This is in good agreement with the 
behavior previously noticed by \citet{lee08} using radio continuum observations 
at 1420~MHz.

The southern radio ridge
is also mimicked by a quite faint optical counterpart. 
Absorption due to the molecular gas mainly located in the foreground of the 
SNR is probably responsible for the observed weakness in the optical emission 
in this part of IC~443.

In the breakout region toward the western side of IC~443
only few local radio enhancements, immersed in  faint diffuse radio 
emission, are observed at the locations of the optical filaments.

\section{Radio spectral properties of SNR~IC~443}	
\subsection{The spatially resolved spectral index of IC~443}
In this section we analyze the spectral
properties of this SNR using the new VLA images presented here at 74 and 
330~MHz. Estimates of the spectral index variations with position across 
the remnant require imaging the interferometric data of IC~443 at both 
frequencies using the same \it uv \rm coverage. 
To perform this, we reconstructed the interferometric images 
by applying appropriate tapering functions to the visibility data at 74 
and 330~MHz. To show more clearly the main spectral features, avoiding any 
masking effect from small scale variations,
we have chosen a final synthesized beam of $70^{\prime\prime}$.
In addition, to avoid any positional offsets, the images were aligned and 
interpolated to identical projections before calculating spectral indices.

\begin{figure*}[!ht]
  \centering
\includegraphics{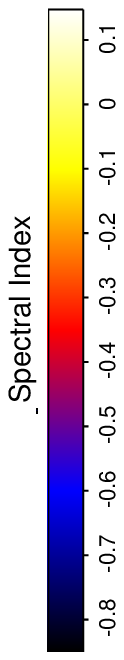}
\includegraphics[width=.65\textwidth]{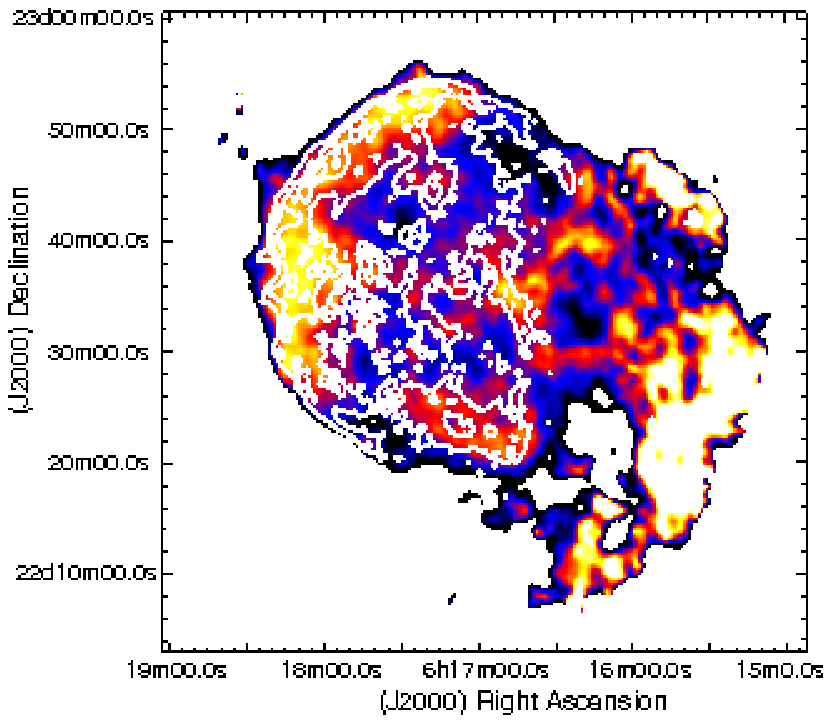}
\caption{Spectral index map constructed using VLA observations at 74 and 
330~MHz (matched to a common angular resolution of $70^{\prime\prime}$). 
The 5.5, 15, 26, and 40 mJy~beam$^{-1}$ contours from the 
330~MHz image are included to facilitate the comparison between spectral 
continuum and total power features. To create this map the 74 and 330~MHz 
images were masked at 3$\sigma$.}
\label{spectralmap-330intensity}
\end{figure*}

The dependency of the spectral index between 74 and 330~MHz with the 
position within IC~443 was determined through 
the construction  
of a spatially resolved spectral index map, which
is shown in Fig.~\ref{spectralmap-330intensity}. 
To produce the spectral map the matched images
of IC~443 at both frequencies were masked at the 3$\sigma$ level of their 
respective noise levels. The error in the determination of the spectral 
index from the map is less than $0.04$ in the high flux regions 
(the east limb of the SNR and some interior filaments) and
about 0.1 in the diffuse central emission. The uncertainties increase
in the weakest emission regions towards the westernmost part of the remnant 
due to the lower sensitivity of the image at 74~MHz in this region (see below).

Figure~\ref{spectralmap-330intensity} shows good agreement between total 
intensity features and the spatial distribution of the spectral index.
The 74/330~MHz spectral map reveals for the first time the morphology of a 
very flat spectral component running  
along the eastern side of IC~443 within which the spectral index varies 
between $\alpha_{74}^{330}\sim-0.05$ and $\sim-0.25$. 
This flattening towards the brightest filaments 
(see also Fig.~\ref{74-subim} and Fig.~\ref{330-subim}) indicates that some thermal
absorption is present, and would become stronger at lower frequencies.
It is remarkable that such spectral behavior 
has also counterparts
in the \it J \rm and  \it H \rm  bands as observed by 2MASS
(further details for the correlation between local spectral index variation 
across IC~443 and the IR emission are described below in Sect.~5).

The large-scale diffuse emission in the SNR's interior has a spectrum that 
is markedly different from the eastern part of the remnant with steeper 
components ranging from $\alpha_{74}^{330}=-0.6$ to a quite steep
value of $\alpha_{74}^{330}=-0.85$, 
as would be expected under the linear diffusive shock acceleration model for
weak shocks with low Mach numbers \citep{and93}.
The southern sinuous ridge, where most of the interaction with the molecular 
cloud is taking place, is seen in our spectral index map as a region with 
$\alpha_{74}^{330}$ varying between $\sim-0.25$ and $\sim-0.5$. 
These local spectral indices are in good correlation with total intensity 
features: the brighter synchrotron areas are systematically
flatter than the other parts of the ridge. 
In this case the interpretation of the flattening in the spectrum is different
from the case noted above in the east rim. Here, the  
spectral behavior might be a signature of Fermi shock 
acceleration at the sites where stronger post compression shock densities,
accompanied by higher local Mach numbers, and/or  higher magnetic field
strength
result due to the impact of the SNR blast
wave on denser ambient medium \citep{bel78,and93}. 

Towards the western side of IC~443, which is dominated by diffuse and faint
radio emission, the spectral indices are mostly steep 
($\alpha_{74}^{330}\simeq-0.60$). Although some flatter features are also 
observed in this region, they are mainly located bordering clipped areas 
in the map and should not be considered as real features owing to the 
decrease in the signal-to-noise ratio observed in this region in the 
VLA~74~MHz image in comparison with the image at 330~MHz.  

The spectrum calculated for the point-like sources near 
R.A.=$06^{\mathrm{h}}\,17^{\mathrm{m}}\,31^{\mathrm{s}}$,
dec.=$+22^{\circ}\,45^{\prime}\,48^{\prime\prime}$ and
$06^{\mathrm{h}}\,17^{\mathrm{m}}\,36^{\mathrm{s}}$, $+22^{\circ}\,24^{\prime}\,57^{\prime\prime}$
confirms their extragalactic nature \citep{bra86b}. 
A similar result is obtained 
for the point radio source located at the position
R.A.$\sim06^{\mathrm{h}}\,15^{\mathrm{m}}\,34.7^{\mathrm{s}}$,
dec.$\sim+22^{\circ}\,41^{\prime}\,46^{\prime\prime}$.

\subsection{The integrated spectrum of IC~443}
To update the mean spectral index determination for IC~443, we have 
included the new integrated flux densities at 74 and 330~MHz in the extensive 
list of measurements presented in the literature. 
In Table~\ref{fluxes} we list the integrated flux density estimates for the SNR
between 10 and 10700~MHz.
We applied a correction factor over the wide spectral range 
between 408 and 10700~MHz in order to place each value on the flux 
scale of \citet{baa77}, except in some cases for which no information was 
available on the flux value considered for the primary calibrators. 

\begin{table*}[h!]
\renewcommand{\arraystretch}{1.0}
\centering
\caption{Integrated flux densities on the SNR~IC443}
\vspace{0.26cm}
{\begin{tabular}[width{22cm}]
{llllll} \hline\hline
Frequency & Scaled flux & References & Frequency & Scaled flux & References \\
(MHz)     & density (Jy)&            &  (MHz)    & density (Jy)         \\ \hline
10\dotfill& $400\pm100^{\mathrm{(a)}}$ & \citet{bri68} & 430\dotfill & 
$245\pm30^{\mathrm{(c)}}$ & \citet{kun68} \\
20\dotfill   & $600\pm144^{\mathrm{(a)}}$ & \citet{bra69} & 513\dotfill &
$205\pm27^{\mathrm{(c)}}$ & \citet{bon65}  \\
22\dotfill   & $615\pm75^{\mathrm{(a)}}$ & \citet{rog86}  & 610\dotfill &
$215\pm32^{\mathrm{(c)}}$ & \citet{dic69}  \\
22.25\dotfill& $535\pm65^{\mathrm{(a)}}$ & \citet{rog69}  & 635\dotfill &
$179\pm18$ & \citet{mil69} \\
22.3\dotfill& $529\pm36^{\mathrm{(a)}}$ & \citet{gui69}  & 740\dotfill &
$164\pm15^{\mathrm{(c)}}$ & \citet{bon65}   \\
25\dotfill   & $630\pm132^{\mathrm{(a)}}$ & \citet{bra69} & 750\dotfill &
$190\pm25^{\mathrm{(c)}}$ & \citet{hog64}    \\
26.3\dotfill& $600\pm48^{\mathrm{(a)}}$  & \citet{vin75} & 960\dotfill &
$196\pm24$& \citet{har60}                      \\
26.7\dotfill& $561\pm33^{\mathrm{(a)}}$  & \citet{gui69}  & 960\dotfill &
$165\pm10^{\mathrm{(c)}}$ & \citet{bon65}       \\
33.5\dotfill& $582\pm37^{\mathrm{(a)}}$  & \citet{gui69}  & 1000\dotfill &
$160\pm16^{\mathrm{(c)}}$ & \citet{mil71}       \\
34.5\dotfill& $440\pm88^{\mathrm{(a)}}$  & \citet{dwa82}  & 1390\dotfill &
$177\pm15$ & \citet{wes58}       \\
38\dotfill   & $650\pm95^{\mathrm{(a)}}$  & \citet{bal54}  & 1400\dotfill &
$170\pm20^{\mathrm{(c)}}$ & \citet{hog64}        \\
38\dotfill   & $730\pm100^{\mathrm{(a)}}$  & \citet{bly57}  & 1400\dotfill &
$146\pm18$ & \citet{wan61}         \\
38\dotfill   & $460\pm46^{\mathrm{(a)}}$  & \citet{wil66} & 1410\dotfill &
$131\pm13$ & \citet{mil69}          \\
38.6\dotfill& $547\pm40^{\mathrm{(a)}}$  & \citet{gui69}  & 1419\dotfill &
$130\pm13$ & \citet{gre86}       \\
74\dotfill&  $470\pm51^{\mathrm{(b)}}$ &  This work & 1420\dotfill &
$160\pm16^{\mathrm{(c)}}$ & \citet{hag55}     \\
81.5\dotfill& $420\pm63^{\mathrm{(a)}}$  & \citet{bal54}  & 1420\dotfill &
$138\pm15$ & \citet{hil72}      \\
81.5\dotfill& $470\pm70^{\mathrm{(a)}}$ & \citet{sha55} & 2650\dotfill &
$86\pm9$ & \citet{mil69}   \\
83\dotfill   & $470\pm80^{\mathrm{(a)}}$ & \citet{kov94} & 2700\dotfill &
$104\pm15$ & \citet{mil71}       \\
102\dotfill  & $480\pm80^{\mathrm{(a)}}$ & \citet{kov94} & 3000\dotfill &
$100\pm15^{\mathrm{(c)}}$  & \citet{hog64} \\
111\dotfill  & $440\pm80^{\mathrm{(a)}}$ & \citet{kov94} & 3125\dotfill &
$100\pm15^{\mathrm{(c)}}$  & \citet{kuz60}  \\
151\dotfill  & $280\pm35^{\mathrm{(a)}}$ & \citet{gre86} & 4170\dotfill &
$100\pm15^{\mathrm{(c)}}$  & \citet{hir72}  \\
159\dotfill  & $270\pm40^{\mathrm{(a)}}$ & \citet{edg59} & 5000\dotfill &
$79\pm11$ & \citet{mil71}  \\
178\dotfill  & $210\pm42^{\mathrm{(a)}}$ & \citet{ben62} & 5000\dotfill &
$85\pm13^{\mathrm{(c)}}$  & \citet{kun69}    \\
195\dotfill  & $290\pm45^{\mathrm{(a)}}$ & \citet{kun68}& 6640\dotfill &
$70\pm15^{\mathrm{(c)}}$  & \citet{dic71}   \\
330\dotfill  & $248\pm15^{\mathrm(b)}$ &  This work & 8000\dotfill &
$90\pm18$ & \citet{how63}   \\
400\dotfill  & $230\pm34^{\mathrm{(a)}}$ &  \citet{dav65} & 10700\dotfill &
$60\pm5^{\mathrm{(c)}}$  & \citet{kun72}  \\
400\dotfill  & $210\pm31^{\mathrm{(a)}}$ &  \citet{sig65} & \\
400\dotfill  & $251\pm8^{\mathrm{(a)}}$ &  \citet{kel64} & \\
408\dotfill  & $289\pm28$ &  \citet{col71}  \\
\hline
\label{fluxes}
\end{tabular}}
\begin{list}{}{}
\item[$^{\mathrm{(a)}}$] No correction to \citet{baa77} scale was applied.
\item[$^{\mathrm{(b)}}$] Flux density scale from VLA Calibrator Manual,
http://www.aoc.nrao.edu/~gtaylor/csource.html.
\item[$^{\mathrm{(c)}}$] The correction factor was not available.
\end{list}
\end{table*}

\begin{figure}[ht]
\centering
\includegraphics[width=8cm]{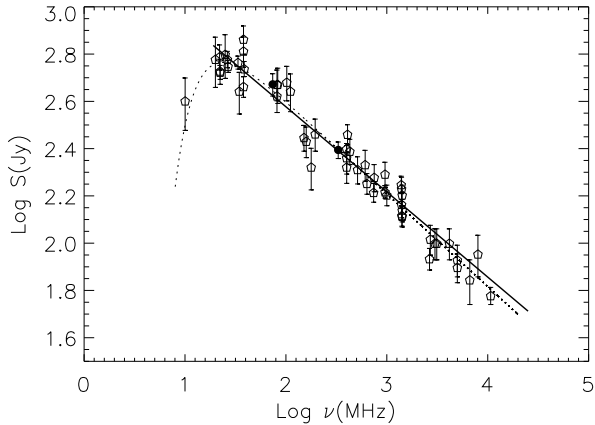}
\caption{Radio continuum spectrum for SNR~IC~443 obtained from the flux
density values listed in Table~\ref{fluxes}. The filled circle 
symbols correspond to the new flux density measurements calculated using 
the VLA data at 74 and 330~MHz presented in this work, the open symbols are 
for radio observations previously published and, where possible, brought 
onto the flux density scale of \citet{baa77}.
Solid line represents the linear fit to the flux density values excluding that at
10~MHz, which produces a spectral index $\alpha= -0.36\pm0.02$ ($S\propto\nu^{\alpha}$).
Dotted line shows a fit to all of the plotted values if absorption were present 
(eq.~\ref{turnover}), which yields a spectral index $\alpha= -0.39\pm0.01$.}
\label{global-spectrum}
\end{figure}

A plot of the integrated radio continuum spectrum for the SNR~IC~443 is 
shown in Fig.~\ref{global-spectrum}. Our new integrated flux measurements at 
74 and 330~MHz are indicated by filled circle symbols.
From the spectrum it is evident that the flux densities measured at the lowest
radio frequencies, and particularly the flux value at 10~MHz, lie below the
general trend of the data
suggesting the presence of thermal
absorption along the line of sight. In order to fix the integrated spectral
index of IC~443 we first 
use a single power law spectrum to fit the spectrum well down to our lowest 
measurement at 74~MHz,
excluding the flux density value at 10~MHz
(represented as a solid line in Fig.~\ref{global-spectrum}). A weighted fit produces a spectral index 
$\alpha=-0.36\pm0.02$ ($S_{\nu}\propto\nu^{\alpha}$).
This result agrees very well within the error
limits with the previous estimates presented by \citet[][ and references
therein]{eri85}. 
If we consider the lowest frequencies observations, these measurements can be fit
with a power law
plus an exponential turnover using the Eq.~\ref{turnover} 
(indicated by the dotted line in Fig.~\ref{global-spectrum}) 

\begin{equation}
S_{\mathrm{\nu}}=S_{\mathrm{330}}\,\left(\frac{\nu}{330\,\mathrm{MHz}}\right)^{\alpha}\,
\mathrm{exp}[-\tau_{330}\,\left(\frac{\nu}{330\,\mathrm{MHz}}\right)^{-2.1}]
\label{turnover}
\end{equation}

Here, 330~MHz is a reference frequency at which an integrated flux density $S_{330}$
and an optical depth $\tau_{330}$ are measured, $\alpha$ represents the non-thermal integrated
spectrum, which is assumed to be constant throughout the radio band. We have made a weighted fit of the distribution of 
data points over four decades in frequency and find for the whole SNR
a single radio spectral index $\alpha=-0.39\pm0.01$,
and an average optical depth $\tau_{330}=(7\pm1)\times10^{-4}$. 
The free-free continuum optical
depth at 10~MHz derived from the relation
$\tau_{10}=\tau_{330}\,[10/330]^{-2.1}$ is $\tau_{10}=1.07$, while the optical
depth at 74~MHz is $\tau_{74}=0.02$. Our results indicate
that although the absorption becomes significantly stronger
at 10~MHz this effect is negligible at 74~MHz.
In addition, we note that the spectral index produced by considering free-free absorption
is consistent with that derived with a power law.
Such a result is not surprising given the results of Sect.~4.1, that clearly 
indicated the presence of thermal absorption, though not at a level to impact the integrated
flux at 74~MHz. Future measurements at intermediate frequencies (e.g. $\sim 30$~MHz) are 
needed to understand if the turnover inferred from the 10~MHz measurement is related to the 
subtle absorption revealed in the spectral index analysis presented in Sect.~4.1. 

\begin{figure}[h]
\centering
\includegraphics[width=8cm]{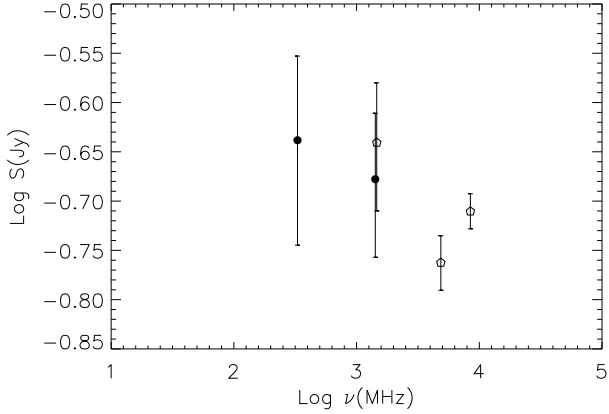}
\caption{Radio continuum spectrum for the pulsar wind nebula (PWN) around
CXOU~J061705.3+222127 in the SNR~IC~443. A weighted fit to the data with a single 
power law function ($S\propto\nu^{\alpha}$) yields a spectral index $\alpha=-0.04\pm0.05$.
The filled circles symbols are from the flux density estimates at 330 and 1420~MHz
presented in the current work, while the open 
symbols are from \citet{olb01}.}
\label{spectrum-pwn}
\end{figure}

Additionally we re-calculate the spectrum of the pulsar wind nebula in IC~443. 
To analyze the global spectrum of the PWN in IC~443 we combine the new flux densities
estimated at 330 and 1420~MHz with data taken from
\citet{olb01} at 1460, 4860, and 8460~MHz. The radio spectrum including all
these measurements is shown in Fig.~\ref{spectrum-pwn}.
The larger error bars for the lower frequency radio data reflect the difficulty in
separating the nebular emission from its surroundings.
A weighted fit to all the flux densities produces a spectral index for the PWN
of $\alpha=-0.04\pm0.05$, which is zero within the uncertainty in the fit itself. This result is
similar to that obtained by \citet{olb01}.

\section{Radio spectral index and the near infrared emission}
\begin{figure*}[!ht]
  \centering
\includegraphics[width=.45\textwidth]{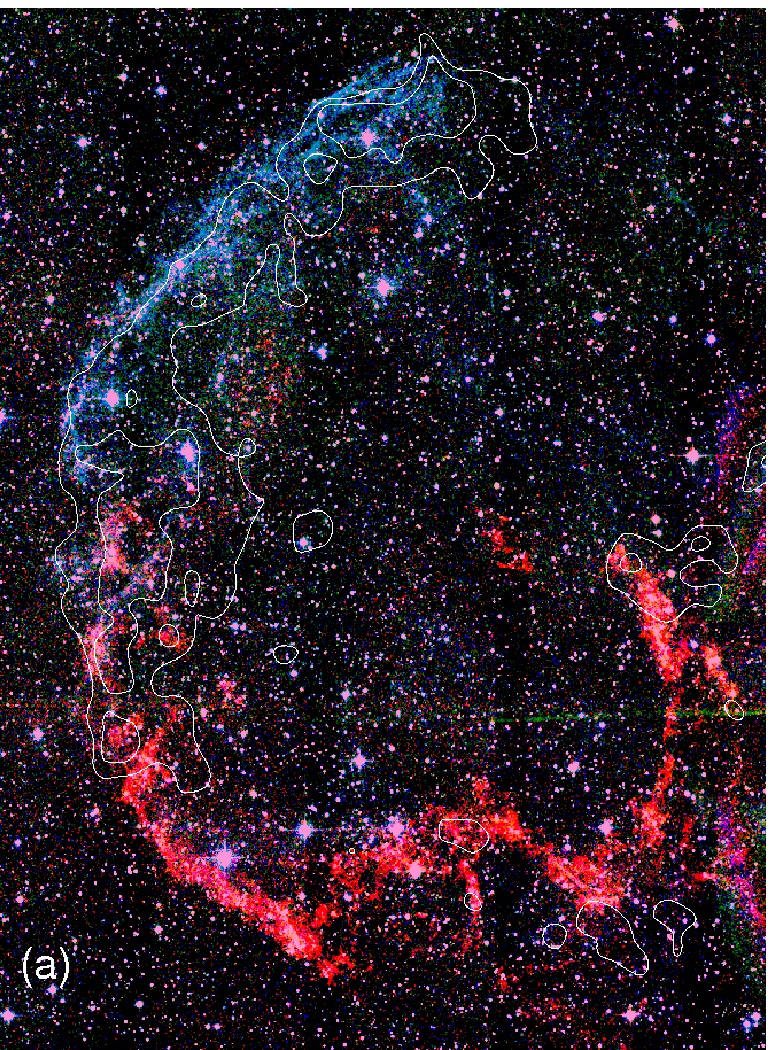}~\hfill
\includegraphics[width=.45\textwidth]{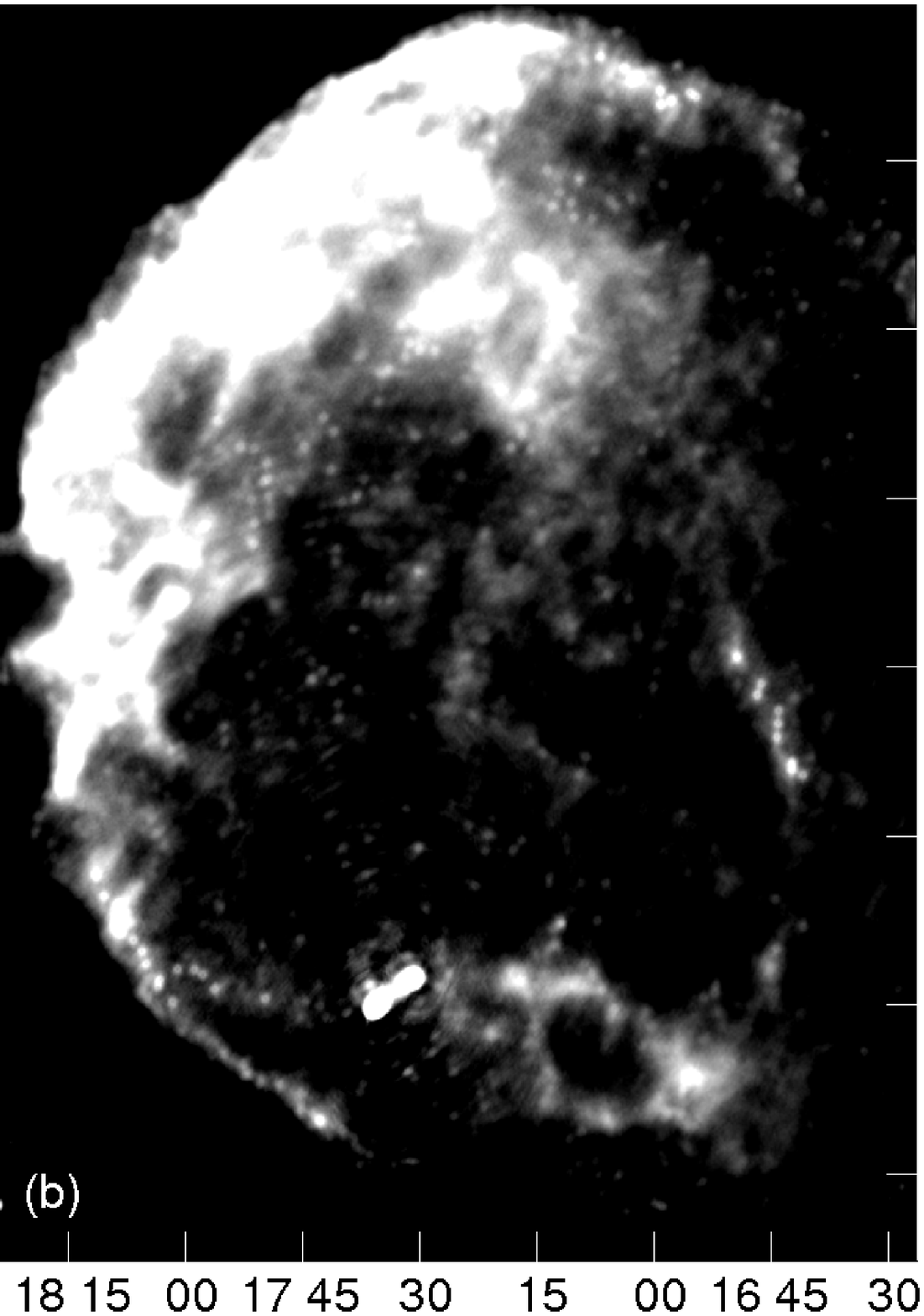}
\caption{\bf (a) \rm A color representation of the near-infrared emission 
observed with 2MASS in the \it J \rm (in blue), \it H \rm (in green), 
and  $K_{s}$ (in red) bands \citep{rho01}. 
The overlaid contours trace the flattest-spectrum radio structures of the 
SNR~IC443 between 74 and 330~MHz at $\alpha_{74}^{330}= -0.05$ and $-0.25$. 
\bf (b) \rm 330~MHz image of IC~443  
showing the locations of the brightest regions of the remnant.
}
\label{ir-radio}
\end{figure*}

Based on our accurate, spatially resolved radio continuum 74/330~MHz spectral map,  
we have investigated the correlation between 
radio spectral features and the near infrared emission (NIR).
Figure~\ref{ir-radio}a shows two 74/330~MHz spectral index contours 
(traced at $\alpha_{74}^{330}=-0.05$ and $\alpha_{74}^{330}=-0.25$) enclosing the east 
rim where we found regions with very flat spectrum in IC~443 
superposed onto the NIR detected in the \it J \rm (1.25~$\mu$m), 
\it H \rm (1.65~$\mu$m), and \it K$_{s}$ \rm (2.17~$\mu$m) bands as taken 
from the Two Micron All Sky Survey (2MASS) \citep{rho01}. 
The 2MASS image of IC~443 shows the dramatic contrast in near infrared color 
between the east rim and the southern portion of the remnant.
In the color representation of the infrared emission, blue traces the 
\it J\rm-band flux, while 
the infrared data in the \it H \rm and \it K$_{s}$ \rm are shown in green 
and red, respectively; white thus enlightens regions where all the three IR 
bands overlap. 
To facilitate the comparison, Fig.~\ref{ir-radio}b displays the 330~MHz 
continuum image of IC~443 with a grayscale selected to emphasize the 
brightest radio emission.

An impressive agreement is observed in location, size and shape
between the NIR emission detected in the \it H \rm and \it J \rm bands 
and the flattest spectral feature as traced by 
the $\alpha_{74}^{330}$ contours along the 
eastern edge of IC~443. This correspondence begins in the northernmost part 
of the remnant and extends down to positions near 
dec.$\sim+22^{\circ}\,25^{\prime}$.
From Fig.~\ref{ir-radio}b it is also notable that the brightest
radio synchrotron emission perfectly matches the bright emission in the NIR 
bands. As noticed by \citet{rho01}, at this site the 
predominant constituent of the emission in the \it J \rm and \it H \rm bands 
is the [FeII] line, with a minor contribution  from other 
multi-ionized species like [NeII], [NeIII], [SiII], [SIII], etc. 
\citet{rho01} proposed a model in which the infrared emission from
the ionized species in the east bright radio limb of IC~443 
comes from shattered dust produced by a fast dissociating J-type shock. 
The present accurate comparison between radio spectral indices and IR emission
confirms this model. In effect, the passage of a dissociative shock through a 
molecular cloud not only dissociates molecules but also ionize the atoms.
Such collisional ionization is responsible for the thermal absorbing electrons
that produce the peculiar very flat spectrum areas observed all along the eastern
border of IC~443. This interpretation  
is in agreement with studies based on CO and X-ray observations which conclude that the large 
molecular cloud complex is located 
in front of IC~443 \citep{cor77,tro06}.
Evidence for a similar situation has also been observed in the
ringlike morphology of 3C~391 \citep{bro05}, another SNR known to be
interacting with a molecular cloud.

Towards the interior of the bright eastern shell, the spectral index 
gradually steepens with position in coincidence with a decrease in the
intensity of the radio emission. Widely distributed $K_{s}$ band emission
is observed in the 2MASS image in this part of the remnant, which was 
proposed to delineate H$_{2}$ shocked gas from the region interacting with 
the adjacent molecular cloud. 

In contrast to the excellent agreement between the ionic emitting gas and the 
flattest spectrum features
found in the eastern bright limb, no much obvious correspondence is
observed in the southern part of IC~443 with the exception of a spectral 
component located at the northern extreme of the ridge, near
R.A.$=06^{\mathrm{h}}\,16^{\mathrm{m}}\,45^{\mathrm{s}}$,
dec.$=22^{\circ}\,35^{\prime}$. This poor IR/radio-spectrum correspondence is
consistent with the hypothesis proposed before, in which the flat spectrum
features in this part of IC~443 have a non-thermal origin.

In the remainder of this section we attempt to infer, using our new measurements at low
radio frequencies, the physical properties of the area of 
thermal absorption seen in Fig.~\ref{ir-radio}a spatially coincident with the
ionized fine-structure line emitting atoms, i.e. the eastern half of the SNR.
At radio wavelengths, the emission measure (EM) is given by,

\begin{equation}
\mathrm{EM}=6.086\times10^{-6}\,\it a(\mathrm{T_{e},\nu)}^{-1}\,\nu^{2.1}\,\tau_{\nu}\,
\mathrm{T_{e}^{1.35}}\, \mathrm{cm^{-6}\,pc},
\label{em}
\end{equation}
 
where \it a\rm($T_{e}$,$\nu$) is the Gaunt    
factor assumed to be 1, a correct value for the
range of astrophysical quantities involved in our calculations;
$\nu$ and $\tau_{\nu}$ are the frequency measured in MHz and the free-free optical
depth, respectively, and $\mathrm{T_{e}}$ is the electron temperature in K of the
intervening ionized gas. By measuring the relative strengths of the 
[FeII] lines observed in the near and mid infrared emission, \citet{rho01}
conclude that the emitting region behind the J-type shocks as observed in the 
\it J \rm and \it H \rm bands of the 2MASS has a temperature of 12000~K. Although
not quantified, the authors recognize a large uncertainty associated with this
magnitude as a consequence of different beam size and possibly different filling
factors in their measurements. 
If we assume that $\mathrm{T_{e}}$ is in a reasonable range between 8000-12000~K 
(which includes the temperature as 
estimated from the IR observations) and use the optical
depth derived from our radio study particularly for this region where the thermal
absorption is stronger ($\tau_{74}\sim0.3$) we obtain an 
EM between approximately $2.8\times10^{3}$ and $5.0\times10^{3}$~cm$^{-6}$~pc 
for the eastern rim.
By combining this emission measure with the postshock electron density, we can 
roughly calculate the thickness of the molecular gas layer that has been dissociated
and ionized by the SNR shock front. If we assume an electron density of 
$n_{e}\sim\,500$~cm$^{-3}$ as estimated by \citet{fes80} and \citet{rea00} on the basis of 
forbidden [FeII] lines, we conclude that the dissociation and ionization processes took place
in a thin screen of about 3.4 to 6.0 $\times10^{16}$~cm ($\sim$0.01-0.02~pc). This is a small path compared 
with the transverse dimensions over which thermal absorption is observed, but is a lower limit
if the ionized gas is clumped.

\section{Comparison with the molecular distribution}
\begin{figure*}[!h]
  \centering
\includegraphics[width=14cm]{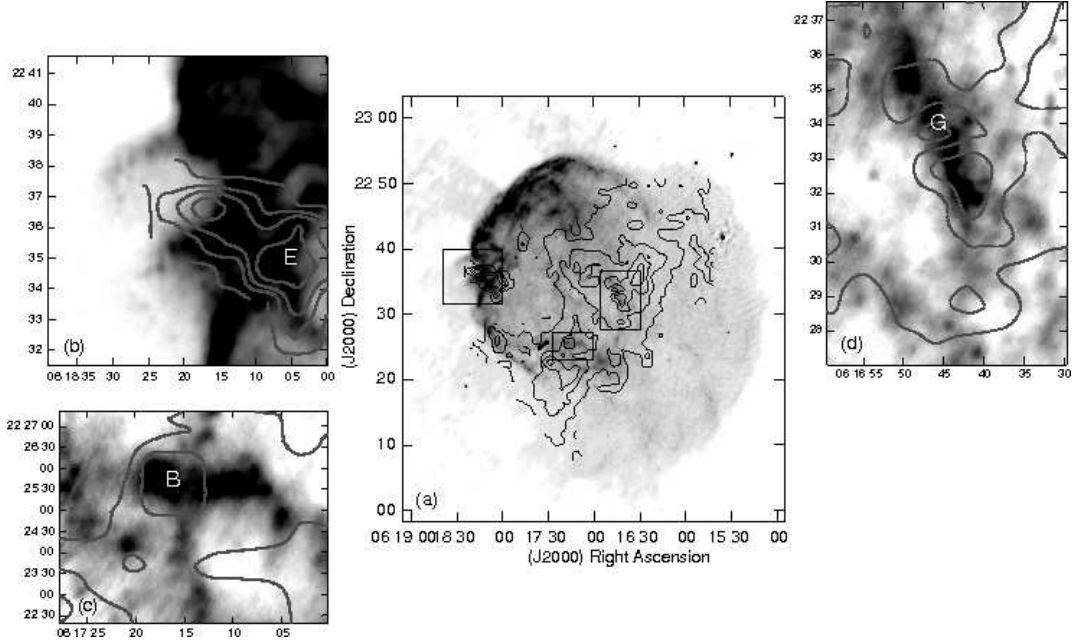}
 \caption{A comparison of the radio continuum emission of 
IC~443 and the $^{12}$CO~($J$=1-0) distribution in the SNR region. 
The grayscale representation corresponds to the new VLA~330~MHz image,
while the superposed contours trace CO integrated emission from $-10$ to 
$-1$~km s$^{-1}$ as taken from \citet{zha09}. Close-ups of three interesting 
areas are displayed around the center image, with the CO contours 
overlapping. The white letter in each panel corresponds to the designation
of the molecular clump by \citet{dic92}.}
\label{co-330}
\end{figure*}

In Fig.~\ref{co-330} we present a comparison between our VLA 330~MHz
image and new $^{12}$CO~($J$=1-0) data presented by \citet{zha09} on the basis
of new observations carried out with the telescope of the Purple Mountain Observatory 
in China (main beam size of $50^{\prime\prime} \times 54^{\prime\prime}$,
velocity resolution of 0.37~km~s$^{-1}$, and rms noise level of 0.1-0.3~K at
a velocity resolution of $\sim$0.2~km~s$^{-1}$).
The contours superposed on the radio emission depict the
CO emission integrated in the range between $-10$
and $-1$~km~s$^{-1}$, which includes the systemic velocity of IC~443.
As described before, the molecular material is preferentially located in the
center of the remnant extending in the southeast-northwest direction.
The spatial distribution of the molecular gas across IC~443 is clearly
non-uniform. Earlier observations have identified the presence of various
clumps of molecular gas with broad line
widths, as expected from the interaction with the supernova shock front
\citep[clumps labeled from A to H in the nomenclature of][]{dic92}.

On the basis of the new image at 330~MHz it is
possible to recognize details previously unnoticed in the spatial
comparison between the radio emission and the molecular gas. 
In Fig.~\ref{co-330} we display the regions where good correlation between
radio features and molecular gas distribution is observed. Particularly
noticeable is Fig.~\ref{co-330}b where 
it is apparent that the indentation of the eastern
border in radio occurs near a region where a significant enhancement in the
CO emission is detected. The molecular complex is transverse to the radio
indentation with the maximum of the CO emission shifted to the southwest 
in at least $3^{\prime}$ from the border of the SNR (in the region of the 
molecular clump E).
It is possible that the singular indentation has formed as the result of the
supernova shock front wrapping around a dense clump.
Also, Fig.~\ref{co-330}c shows the presence of a concentration in the CO
emission around
R.A.$= 06^{\mathrm{h}}\,17^{\mathrm{m}}\,16^{\mathrm{s}}$,
dec.$= +22^{\circ}\,25^{\prime}\,40^{\prime\prime}$
(molecular clump B), that matches a local maximum in the radio
emission.
The morphological matching between the radio synchrotron emission
and the molecular gas is especially remarkable at the northern extreme of
the southern ridge of IC~443 near
R.A.$= 06^{\mathrm{h}}\,16^{\mathrm{m}}\,45^{\mathrm{s}}$,
dec.$= +22^{\circ}\,34^{\prime}\,00^{\prime\prime}$
(as shown in Fig.~\ref{co-330}d), where the molecular contours, delineating
high density gas in the region of the clump G, are observed enclosing the
bright radio emission.

\begin{figure}[!ht]
  \centering
\includegraphics[width=8cm]{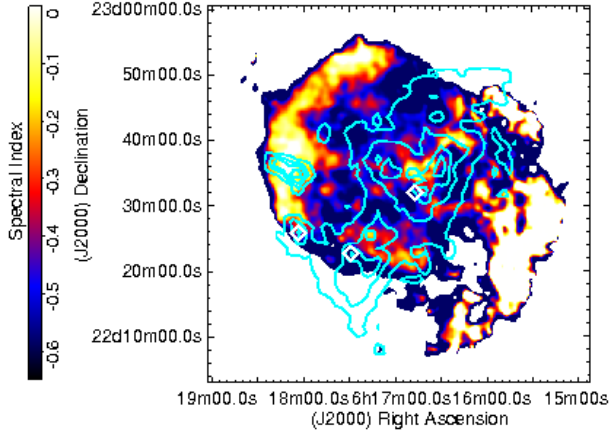}
 \caption{A comparison of the spectral index distribution  with
the molecular emission. The color representation corresponds to the spatially
resolved spectral index map between 74 and 330~MHz, while the overlaid
contours trace the $^{12}$C0~($J$=1-0) emission integrated between
$-10$ and $-1$~km~s$^{-1}$ \citep{zha09}. The positions
where OH(1720~MHz) maser emission were detected are indicated with open white
diamonds \citep{hof03,hew09}.}
\label{co-alpha}
\end{figure}

We also searched for spectral evidence of shock/cloud interaction. Figure~\ref{co-alpha} 
displays an overlay of the 
$^{12}$CO~($J$=1-0) integrated emission contours with the radio spectral
index distribution calculated between 74 and 330~MHz. The CO molecular gas seen in projection
onto the plane of the sky, overlaps the only flat spectral region 
observed in the interior of IC~443. Various small components with a spectrum appreciably 
flatter than the surrounding synchrotron plasma are observed distributed nearby or in 
coincidence with local higher density gas as traced by the $^{12}$CO contours 
along the bright southern ridge and towards the northwest, suggesting that these 
features must be regions where strong shocks encountered denser material 
\citep[as discussed by][]{and93}. In the eastern periphery the situation is different.
The very flat spectrum component extends over an area considerably larger than 
the molecular cloud, precisely because in this region, as shown in Sect.~5, most of
the molecules were dissociated and ionized, absorbing the radio emission at low
frequencies.

In Fig.~\ref{co-alpha} we have identified with open white diamonds the regions with 
OH~(1720~MHz) maser emission as detected by \citet{hew09}.
In the case of SNR~W28, \citet{dub00} demonstrated a clear correspondence
between regions of flat spectral index and OH maser emission.
We search for a similar correlation in IC~443. From Fig.~\ref{co-alpha} we conclude 
that there is not a simple association
between maser locations and spectral index features but rather both flat and steep
components are observed near masers areas.
This fact can be compared with previous results, which suggest that
maser regions in IC~443 arise from regions with
different shock geometry: a shock mostly propagating
towards the line of sight in the
southern OH maser emission and a transverse shock in the
westernmost OH emitting region \citep{cla97,hof03,hew09}.
We find that the OH maser area with transverse shock
(R.A.$=06^{\mathrm{h}}\,16^{\mathrm{m}}\,43^{\mathrm{s}}.6$,
dec.$=+22^{\circ}\,32^{\prime}\,36^{\prime\prime}.7$)
correlates well with flat spectral index ($\alpha_{74}^{330}=-0.4$), 
and one of the OH regions where the shock is tangential to the line of sight
(R.A.$=06^{\mathrm{h}}\,17^{\mathrm{m}}\,29^{\mathrm{s}}.29$,
dec.$=+22^{\circ}\,22^{\prime}\,42^{\prime\prime}.5$)
is associated with a steep spectrum ($\alpha_{74}^{330}=-0.8$).
The fact that flat spectrum emission ($\alpha_{74}^{330}=-0.4$) is 
observed towards the easternmost OH maser area with tangential shock 
(R.A.$=06^{\mathrm{h}}\,18^{\mathrm{m}}\,3^{\mathrm{s}}.67$,
dec.$=+22^{\circ}\,25^{\prime}\,53^{\prime\prime}.4$) 
can be explained because in this region thermal electrons mask the 
intrinsic SNR spectral effects.

\section{Comparison with TeV emission}
\begin{figure*}[!ht]
  \centering
\includegraphics[]{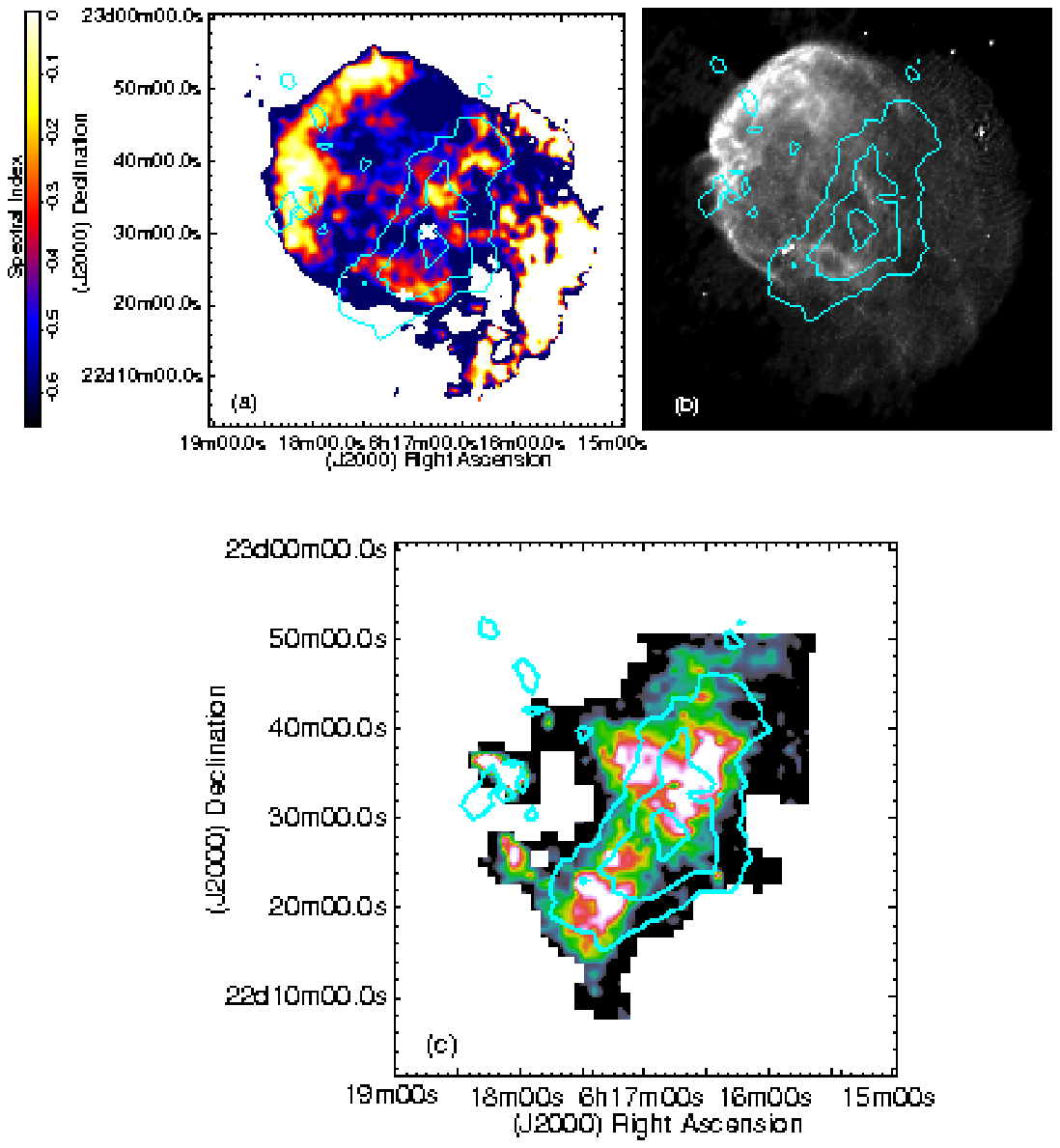}
\caption{Very high energy gamma-rays contours as taken from 
\it VERITAS \rm \citep{acc09} superposed with \bf (a) \rm the 74/330~MHz
spectral index map presented in Fig.~\ref{spectralmap-330intensity}, 
\bf (b) \rm radio continuum emission at 330~MHz, and 
\bf (c) \rm $^{12}$C0~($J$=1-0) integrated emission from $-10$ to $-1$
km~s$^{-1}$ tracing the molecular cloud interaction with IC~443 \citep{zha09}. 
The white plus sign and the white cross in \bf (a) \rm mark the positions 
of the source CXOU~J061705.3+222127 and the centroid of the TeV source
VER~J0616.9+2230, respectively.}
\label{all-gamma}
\end{figure*}

Figure~\ref{all-gamma} compares the TeV gamma-ray significance contours
obtained from \it VERITAS \rm observations \citep{acc09} along with 
the 74/330~MHz spectral index map presented in 
Fig.~\ref{spectralmap-330intensity} (Fig.~\ref{all-gamma}a), with the total
intensity features of IC~443 at 330~MHz (Fig.~\ref{all-gamma}b), and with 
the $^{12}$CO~($J$=1-0) molecular gas distribution (Fig.~\ref{all-gamma}c).
In Fig.~\ref{all-gamma}a the white plus sign marks the position of the source
CXOU~J061705.3+222127, while the white cross indicates the centroid of the
TeV source \object{VER~J0616.9+2230}.

Very high energy (VHE) 
gamma-ray emission roughly extends along the southeast-northwest axis, 
overlapping most of the central region of the remnant.
On the basis of the available statistics, no correlation can be demonstrated
between the radio spectrum or radio features with the TeV emission
region. 
On the contrary, it is remarkable the morphological coincidence between the TeV
emission and the molecular gas distribution (Fig.~\ref{all-gamma}c), at least
up to the level that the available TeV statistic permits to confirm. The interpretation
of this striking correspondence is beyond the scope of this paper.

\section{Conclusions}
In this work we report on new full-synthesis imaging of the Galactic
SNR~IC~443 generated from multiple-configuration
VLA observations at 74 and 330~MHz. 
These high fidelity images constitute the best angular resolution, low frequency radio study
published to date on this classic remnant. Based on these new data
we measured integrated flux densities for this object
of S$_{\mathrm{74\,MHz}}=470\pm51$~Jy and S$_{\mathrm{330\,MHz}}=248\pm15$~Jy. 
On the basis of these new total flux density measurements at 74 and 330~MHz together 
with previously published values, we recalculated the integrated spectrum of IC~443.
We have made this analysis, for the first time, taking into account that
the spectrum of IC~443 has a reported turnover at the lowest radio frequencies.
The fit over a wide frequency range produces a radio spectral index
$\alpha_{10\mathrm{MHz}}^{10700\mathrm{MHz}}=-0.39\pm0.01$ with free-free thermal optical 
depth $\tau_{330}\sim7\times10^{-4}$ (or $\tau_{74}\sim0.02$). 
If the measurement at 10~MHz is excluded, we note that $\alpha$ is relatively
unchanged from that inferred from a single power law fit 
($\alpha_{10\mathrm{MHz}}^{10700\mathrm{MHz}}=-0.36\pm0.02$).  

Based on the combination of the new images at 74 and 330~MHz, we
investigated spectral changes with position across IC~443 and related them
to the spatial characteristics of the radio continuum emission and of the surrounding interstellar
medium. For the first time we spatially resolved the flattest spectrum region
over the SNR (with very flat indices down to $\alpha_{74}^{330}\sim-0.05$)
along the brightest rim on the eastern side of the remnant. The radio spectrum
steepens towards the more diffuse component in the
interior of IC~443 with $\alpha_{74}^{330}$ between $-0.6$ and $-0.85$,
consistent with those expected from linear diffusive shock
acceleration processes. Finally, another region with non-uniform flattening in
the spectrum is identified near the southern ridge of IC~443. We investigated the connection between 
these spatial spectral variations and IR and molecular
emission distribution. 

The comparison with 2MASS NIR data
underscored an impressive coincidence between the eastern radio flattest
spectrum region and NIR ionic lines which account
for most of the infrared emission observed towards this part of IC~443. Based on the presence
of this IR emission, which confirms the existence of a strong J-type shock 
dissociating molecules and ionizing atoms, we conclude that the 
most likely explanation for the flattest spectrum observed here, is that it is
produced by free-free absorption at 74~MHz along the line of sight (with $\tau_{74}$ up to
$\sim 0.3$). From the physical parameters of the IR emitting gas together with our
new radio spectral study, we conclude that the observed thermal absorption
takes place in a very thin layer (width $\sim0.02$~pc)
that extends all along the eastern border.
Such a thin layer would be the product of the
dissociating/ionizing action of the SNR shock over the adjacent molecular gas. It is important to
remark that the low optical depths derived for the whole SNR ($\tau_{330}\sim7\times10^{-4}$ and
$\tau_{74}\sim0.02$) indicate that the 74 and 330~MHz integrated emission
is not significantly attenuated, highlighting that the thermal absorption
inferred from the spatially resolved spectral index map is a relatively subtle and localized effect. 
Our result represents only
the second case, following 3C~391 \citep{bro05}, of spatially
resolved thermal absorption delineating the interaction of a SNR/molecular
cloud shock boundary. Moreover it confirms such phenomena as common and a
rich area of investigation for future, low frequency studies of Galactic complexes.

On the other hand, from the molecular studies it is known
that the southern ridge, where the second region with flat spectrum was identified, 
is the site in which the most complex interaction between the SNR shock
and the external molecular cloud is occurring. 
We used Zhang et al. (2010)'s observations of the $^{12}$CO~($J$=1-0)
molecular gas towards IC~443
to analyze correspondences between
these data and our new low radio frequency observations.
The correlation of the emission at 330~MHz with the new $^{12}$CO
observations revealed spatial irregularities in the denser molecular gas
that are well associated with features of bright emission in radio and  
good agreement between the flat radio spectrum region towards the south 
and molecular emission. We conclude that here the flat radio spectrum  is predominantly a
signature of shock acceleration in a region with strong postshock densities
and enhanced magnetic fields produced after the interaction of the blast wave with
dense ambient medium. 

Furthermore, from the comparison of the molecular environment mapped
by $^{12}$CO data with the local variations of the spectral index,
we find evidence that the shocked molecular gas, as illuminated by the
OH (1720~MHz) maser emission, is coupled with a flattening in the radio
spectral index in the locations where the shock is transverse to the line of sight,
while a steep spectrum is observed in a OH maser region in which
the shocks are propagating along the line of sight. \rm

From the new images, we also analyzed the pulsar wind
nebula around the source CXOU~J061705.3+222127 in IC~443.
Based on the new VLA image at 330~MHz we have derived a flux density
$S_{\mathrm{330\,MHz}}^{\mathrm{PWN}}=0.23\pm0.05$. We have also
measured the flux density of the PWN in the radio continuum image at
1420~MHz presented by \citet{lee08} obtaining
$S_{\mathrm{1420\,MHz}}^{\mathrm{PWN}}=0.20\pm0.04$.
From the combination of these data and measurements taken from the literature we derived
an integrated radio spectrum
$\alpha^{8460}_{330}\sim0.0$ for the PWN.

In addition, on the basis of available \it VERITAS \rm statistics we compared the 
TeV emission with the $^{12}$CO distribution, finding an excellent
morphological correlation between the high energy emission and the
distribution of the molecular gas.
No correspondence was found, however, between the gamma-ray emission as observed
by \it VERITAS \rm and radio spectral or intensity features in radio. A future paper
will address these findings in connection with 
the origin of the gamma-ray emission detected in IC~443.

\acknowledgements{We are very grateful to T.~B. Humensky for kindly
providing us the \it VERITAS \rm significance contours,
to Z. Zhang for the $^{12}$CO~($J$=1-0) data, and to B-C.~Koo for the
radio continuum image at 1420~MHz. We acknowledge the very useful comments of the
anonymous referee.
This publication makes use of data products from the
Two Micron All Sky Survey, which is a joint project of the University of
Massachusetts and the Infrared Processing and Analysis Center/California
Institute of Technology, funded by the National Aeronautics and Space
Administration and the National Science Foundation.
The optical image used in this work is from the Second Palomar Observatory
Sky Survey (POSS-II), which was made by the California Institute of
Technology with funds from the National Science Foundation, the National
Geographic Society, the Sloan Foundation, the Samuel Oschin Foundation,
and the Eastman Kodak Corporation.
This research has made use of the NASA's ADS Bibliographic Services.
Data processing was carried out using the HOPE~PC cluster at IAFE.
This research was partially funded through CONICET (Argentina) grant PIP~112-200801-02166,
ANPCYT-PICT (Argentina) grant 0902/07, and ANPCYT-PICT (Argentina) 08-0795 grant.
Basic research in radio astronomy at the Naval Research Laboratory is supported 
by 6.1 base funds.
}

\bibliographystyle{aa}
\bibliography{paper-ic443}
\IfFileExists{\jobname.bbl}{}
{\typeout{}
\typeout{****************************************************}
\typeout{****************************************************}
\typeout{** Please run "bibtex \jobname" to optain}
\typeout{** the bibliography and then re-run LaTeX}
\typeout{** twice to fix the references!}
\typeout{****************************************************}
\typeout{****************************************************}
\typeout{}
}

\end{document}